\documentclass[preprintnumbers,superscriptaddress,reprint]{revtex4-1}
\usepackage{textcomp}
\usepackage{amsmath}
\usepackage{amssymb}
\usepackage{graphicx}
\usepackage{color}
\usepackage{soul}
\newcommand{\cmp}
{\affiliation{Condensed Matter Physics Division, 
Saha Institute of Nuclear Physics, 1/AF Bidhannagar, Kolkata 700064, India.}}
\newcommand{\barasat}
{\affiliation{Barasat Government College, Barasat, Kolkata 700124, India.}}
\newcommand{\Germany}
{\affiliation{Max Planck Institute for Dynamics and Self-Organization, Am Fassberg 17, 37077 G\"{o}ttingen, Germany}}

\begin{document}

\title{Record breaking statistics near second order phase transitions}

\author{Mily Kundu}
 \email{mily.kundu@saha.ac.in}
 \cmp
\author{Sudip Mukherjee}
 \email{sudip.mukherjee@saha.ac.in}
 \barasat \cmp
 \author{Soumyajyoti Biswas∗}
 \email{soumyajyoti.biswas@ds.mpg.de}
 \Germany
 
 \date{\today}

 \begin{abstract}
When a quantity reaches a value higher (or lower) than its value at any time before, it is said to have made a record. 
 We numerically study the statistical properties of records in the time series of order parameters in different models near their critical points.
Specifically, we choose transversely driven Edwards-Wilkinson model for
interface depinning in (1+1) dimensions and the Ising model in two dimensions, as paradigmatic and simple examples of non-equilibrium and equilibrium
critical behaviors respectively. The total number of record breaking events in the 
time series of the order parameters of the models show maxima when the system is near criticality. The number of record breaking 
events and associated quantities, such as the distribution of the waiting time between successive record events, show power law scaling near the critical point. The exponent values 
are specific to the universality 
classes of the respective models. Such behaviors near criticality can be used as a precursor to imminent criticality i.e. abrupt and catastrophic changes in the system.
Due to the extreme nature of the records, its measurements are relatively free of detection errors and thus provide a clear signal regarding the 
state of the system in which they are measured.

\end{abstract}
\maketitle

\section{Introduction}
A record breaking event is the extreme value of any quantity that has assumed the highest or the lowest value of that quantity up to that time. 
Records are always exhilarating irrespective of the fields they are associated with. In case of sports, for example, 
people can very easily recall the name of the highest goal scorer or the names of the gold medal winners in 
Olympic events. Records can also attract attention even when they are linked with disasters such as the largest earthquake of a region, the 
widest spread of an epidemic or the warmest year in recorded history  etc.

Recently, systematic studies of record statistics~\cite{sudip_sm1,sudip_sm2,sudip_sm3} have gained prominence in statistical physics. Quantities such as 
the number of record breaking events with time, waiting time between successive records, can shed light on the emergent correlation
in the underlying system. These quantities~\cite{Tata, Glick} can be evaluated exactly if the time series 
of the quantity measured are  identical independently distributed (i.i.d)~\cite{sudip_van}
events (see \cite{rev} for a review) i.e. if the events were temporaly uncorrelated. Under that condition, the number of record breaking events 
are independent of the probability distribution from which the sizes of the events are drawn. For this reason, if the number of record breaking events
do not follow what is expected from an uncorrelated time series, this would signal the existence of a temporal correlation in the events. 

The advantage of studying record statistics, over other temporal signals, is the fact that it is an absolute and unambiguous quantity. It simply
notes the events that are either the largest or the smallest value up to that time and does not concern with the magnitude by which it is the largest
or the smallest. While measuring the largest events, it is insensitive to small scale measuring 
limitations that can come from the measuring probes, from external noise (e.g. fluctuating 
temperature etc.) or combinations thereof. A record breaking event, being the strongest signal till that time, is unlikely to be
corrupted by the limitations mentioned above. Therefore, studies in record 
statistics gained recent prominence in climate science research \cite{redner06} (detecting global warming through record breaking
temperature). They are also important
in other diverse scientific areas of research, such as 
evolutionary and cellular biology~\cite{sudip_Kauffman,sudip_anderson,sudip_krug,sudip_franke}, earthquake time series \cite{yoder10,davidsen08}, driven disordered systems in general \cite{sch13}, financial data~\cite{sudip_wergen,sudip_wergen2}, spin glass systems~\cite{sudip_jensen,sudip_sibani,sudip_sibani2}, creep rupture events prior to breakdown of materials \cite{danku14} and so on.

A change in the statistics of the record numbers from what is expected from an uncorrelated time series always signals an underlying change in the process with which
the time series is associated. For example, a deviation in the record numbers in sports can signal a development in understanding or change of rule in that sport, in
fracture and breakdown processes this signals an imminent catastrophic events. Therefore, a systematic study of the record statistics at the onset of correlation 
of any system can help in predicting imminent changes in that system.

In this work we study the record statistics near a critical point. The critical point of a system is  where the 
correlation  is spread across the whole system. As the system is far from showing random response, the deviation in the record statistics from the
random statistics is expected to be maximum near the critical point. Many of the physical systems mentioned 
above can undergo drastic and often catastrophic changes across their respective critical points. Therefore, a precursor from
the record statistics can be a useful way to capture such a catastrophic change in a way that is largely not influenced by external noise or
measuring inaccuracies.

In particular, we look into two prototype models in equilibrium and non-equilibrium phase transitions, viz the Ising model
in two dimensions and the Edwards-Wilkinson (EW) model in (1+1) dimensions. The choices are guided by the simplicity and universal
applicabilities of the models. The Ising model is a model with nearest neighbor ferromagnetic interactions among spins with up/down
symmetry placed in a lattice. Depending upon the external temperature, the system can undergo a phase transition from a fully aligned 
ferromagnetic state to a randomly oriented paramagnetic state. The Ising model is a generic example for the equilibrium phase transition that has applications
in a very wide range of systems where such transitions are observed \cite{ising_rmp}, e.g. magnetism, binary solids, neural networks, sociophysics models
to  name a few. The EW model is an elastic manifold driven through a higher dimensional disordered medium, for example a one dimensional elastic line driven through a two dimensional 
medium with quenched disorders or pinning centers. Depending on the strength of the pinning forces and the external drive, the elastic line can be pinned or move 
with a steady velocity in the long time. This is an example of non-equilibrium depinning transition, 
that is both the
simplest (in terms of elastic nature of the manifold) and widely applicable (equivalence with Burridge-Knopoff
model of earthquakes \cite{biswas13} etc.). We take these two simple models to study the behavior of the record statistics near their respective
critical points. Quantification of the behavior of the record statistics and their association with the universality classes of the two models
is the main aim of this work. 

We measure the record statistics in the time series of the respective order parameters i.e. velocity of the line for the
EW model and magnetization per spin for the Ising model.  In general, we find that the number of record breaking events is maximum at the
critical point. Furthermore, the growth of the number of record breaking events with time at the critical point shows a power law
behavior, as does the waiting time distribution of the events. The corresponding exponents are characteristics of the universality class involved
in the critical behavior.
Far away from the critical point, the correlation  in the time-series vanishes and the record number returns to the i.i.d statistics i.e. assumes the values expected from 
uncorrelated time series.

The remaining part of the paper is arranged as follows. In section II we evaluate the 
pinning-depinning transition point of EW interface having uniform distribution of pinning force, using Monte Carlo simulation. 
The numerical studies on interface velocity shows that the variation of record number with time (at the critical point) follows a growing power law in the asymptotic time limit.
Such numerical study also extract the nature of waiting time distribution of observing successive records, revealing a power law fall with increasing waiting time. Those power law exponents are expected to be universal (like critical exponents). Such universality 
is confirmed through the study of the same model with the Gaussian distribution of pinning force. 
In section III we perform Monte Carlo study on 2d-Ising model (nearest neighbor) to extract the variation of 
record number with time. We numerically evaluate the distribution function of waiting time. 
To check the universality in the nature of record statistics we  
repeat these studies for the same model with next nearest neighbor interaction. 

\section{Record statistics of one dimensional Edwards-Wilkinson interface}
Propagation of an interface through a disordered medium is a very common situation arising in various 
branches in physics, e.g. flux lines in type-II superconductors \cite{larkin79}, magnetic domain walls \cite{zapperi98}, 
charge density waves \cite{fisher85}, wetting front \cite{chevalier15},
fracture front \cite{bouchaud93} and so on. The front often represents the interface between two different states in the material, e.g. 
up and down spins in magnetic systems, broken and intact parts of a solid in case of fracture and so on.
 The `elastic' nature of the front depends on the particular physical
system, e.g. in fracture it is often taken as a $1/r^2$ type interaction following linear elastic fracture mechanics, where $r$ is the distance between 
the location of perturbation of the stress field and the point where the perturbation is measured. 

The two competing forces in the phase transition are the externally applied force on the front that drives it towards 
a propagating state with constant velocity and the randomly placed pinning centers that prevent such a propagation.
Given a configuration for the pinning forces, the interface starts propagating beyond a critical value of the external force,
hence the transition. In case of magnetic domain walls, the external force can be an applied magnetic field and the pinning centers can be 
impurities in the material. For fracture (mode-I), the external force is the transverse force applied on the material and the 
pinning centers can be different fracture strengths within the sample, and so on. 
The associated intermittent dynamics of the elastic line shows signatures of critical behavior,
which is determined by the range of interaction of the interface, given a uncorrelated disorder distribution.

Among the various nature of the interaction of the elastic front, nearest neighbor linear elastic interaction is the
simplest that gives non-trivial transition. This is called the Edwards-Wilkinson (EW) interface and can mathematically be 
expressed as \cite{kardar}
\begin{eqnarray}
  \frac{\partial h(x,t)}{\partial t}&=&\nu \triangledown^2 h+\eta (x,h)+F_{ext}\label{1}
\label{ew_eq}
\end{eqnarray}
where $h(x,t)$ represents the height of the interface (measured from some arbitrarily set level) at position $x$ at time $t$, $\nu$ is a constant related to the dynamics surface tension and taken as 1 here,  $\eta$ is a quenched noise representing the disordered medium, and $F_{ext}$ denotes the externally applied force. When an external force is applied to an interface in presence of quenched noise, the motion of the interface shows a depinning transition (i.e. the interface starts moving with a constant velocity) depending on the magnitude of the external force, $F_{ext}$. The interface is pinned i.e. the interface stops moving after a certain time if $F_{ext}$ is weak compared to the quenched noise.  At a  critical external force $F_c$, the interface undergoes a pinning-depinning transition.

The critical behavior of the model is well studied in various contexts \cite{Jean,Stefano}. Here, however, we will focus on the record breaking events on the time series of the order parameters near the critical point. 
\begin{figure}[ht]
\centering
\includegraphics[width=8cm]{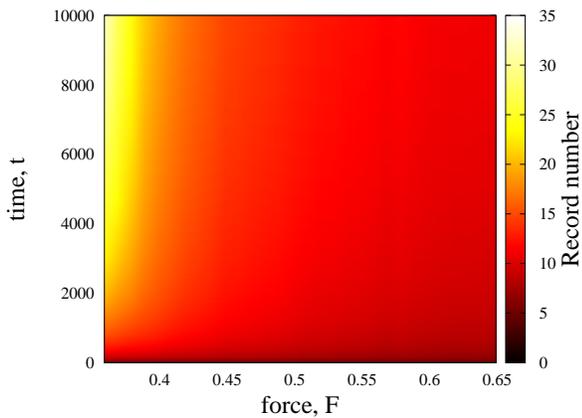}
\caption{For a given external force, the number of record breaking (highest) values for the velocity of the EW interface is shown at different times. The measurements are done after the system reached the steady state. Therefore, only the measurements after the depinning transition can be taken. The value of the number of records is the highest near the critical point and decreases monotonically away from the critical point.}
\label{EW_den}
\end{figure}
\begin{figure}[ht]
\centering
\includegraphics[width=8cm]{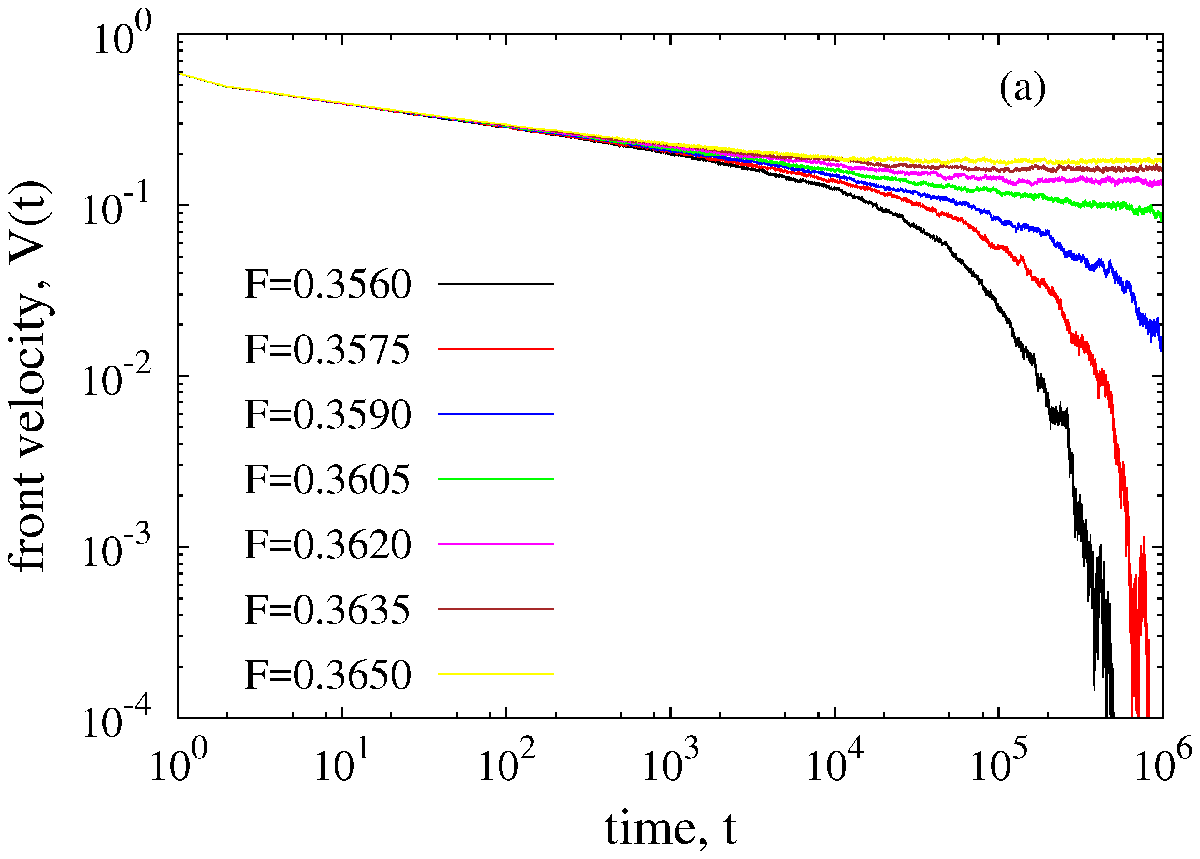}
\includegraphics[width=8cm]{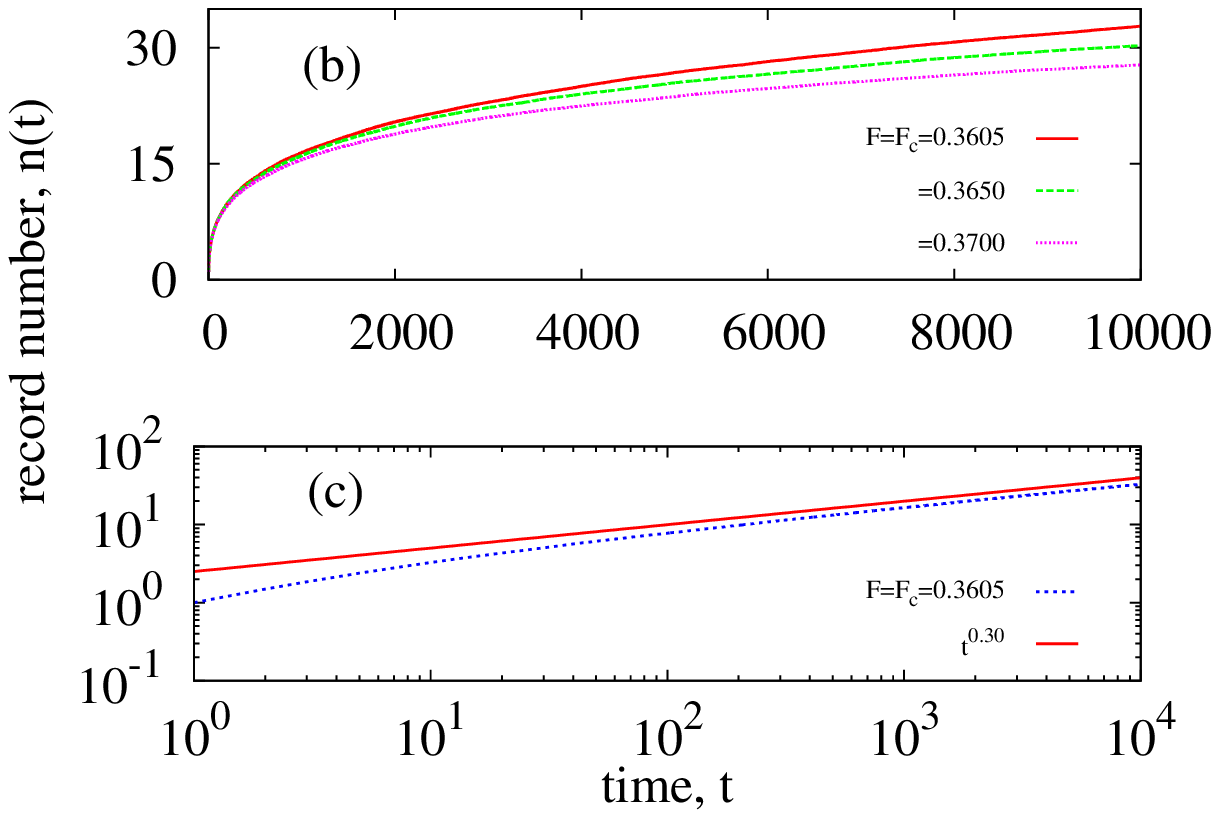}
\caption{The behaviors of the order parameter (velocity) and number of record breaking events in the EW model near the depinning transition point for system size $N=10^5$. (a)Variation of velocity ($V$) with time of EW interface for different driven force $F_{ext}$. Steady state velocity is achieved for $F_{ext}=0.3605=F_c$. (b)Variation of record number $n(t)$ with time $t$ for different values of $F_{ext}$. The curve for $F_{ext}=0.3605$ supersedes the other plots indicating the critical force $F_c=0.3605$, which is the
same estimate obtained from the velocity as well. Part (c) shows a power law fitting of $n(t)$ in large $t$ limit, gives the exponent $\alpha^{EW}_{n} = 0.30 \pm 0.01$.}
\label{velocity_record}
\end{figure}

\begin{figure}[ht]
\centering
\includegraphics[width=8cm]{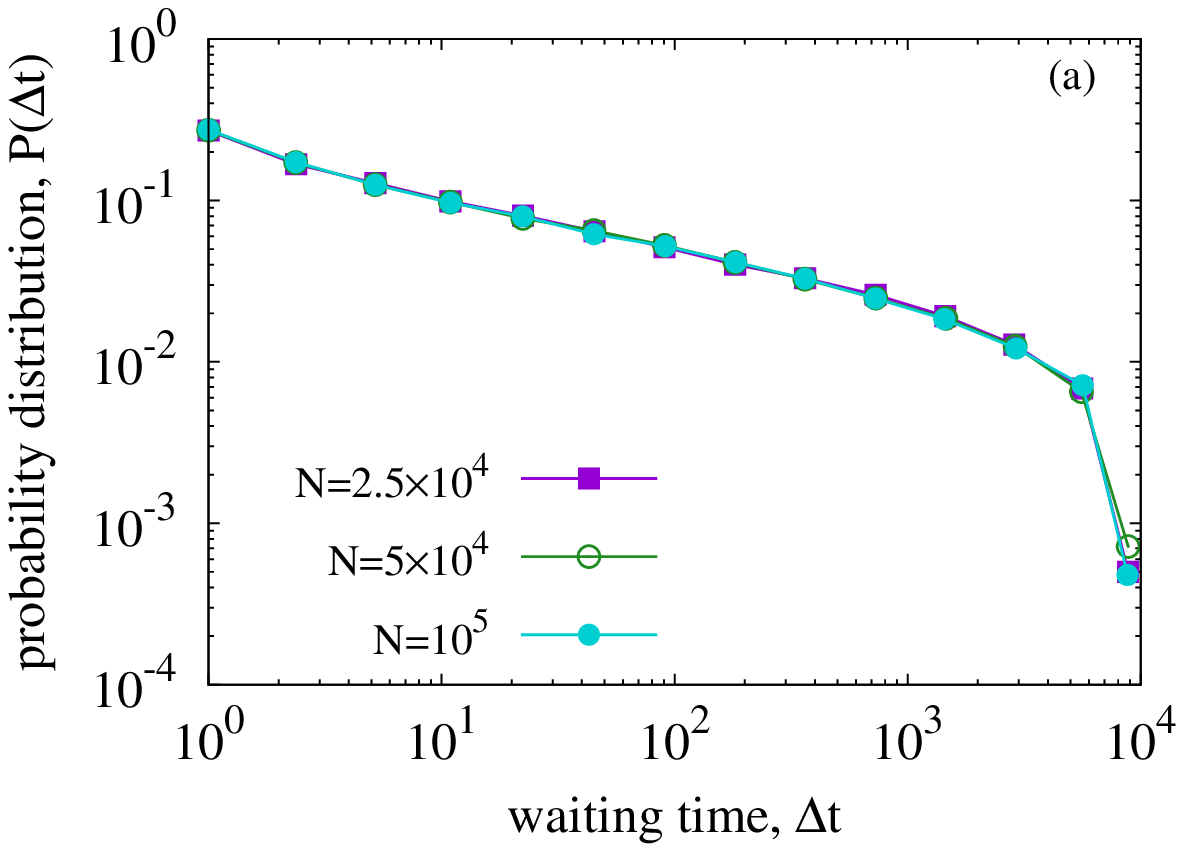}
\includegraphics[width=8cm]{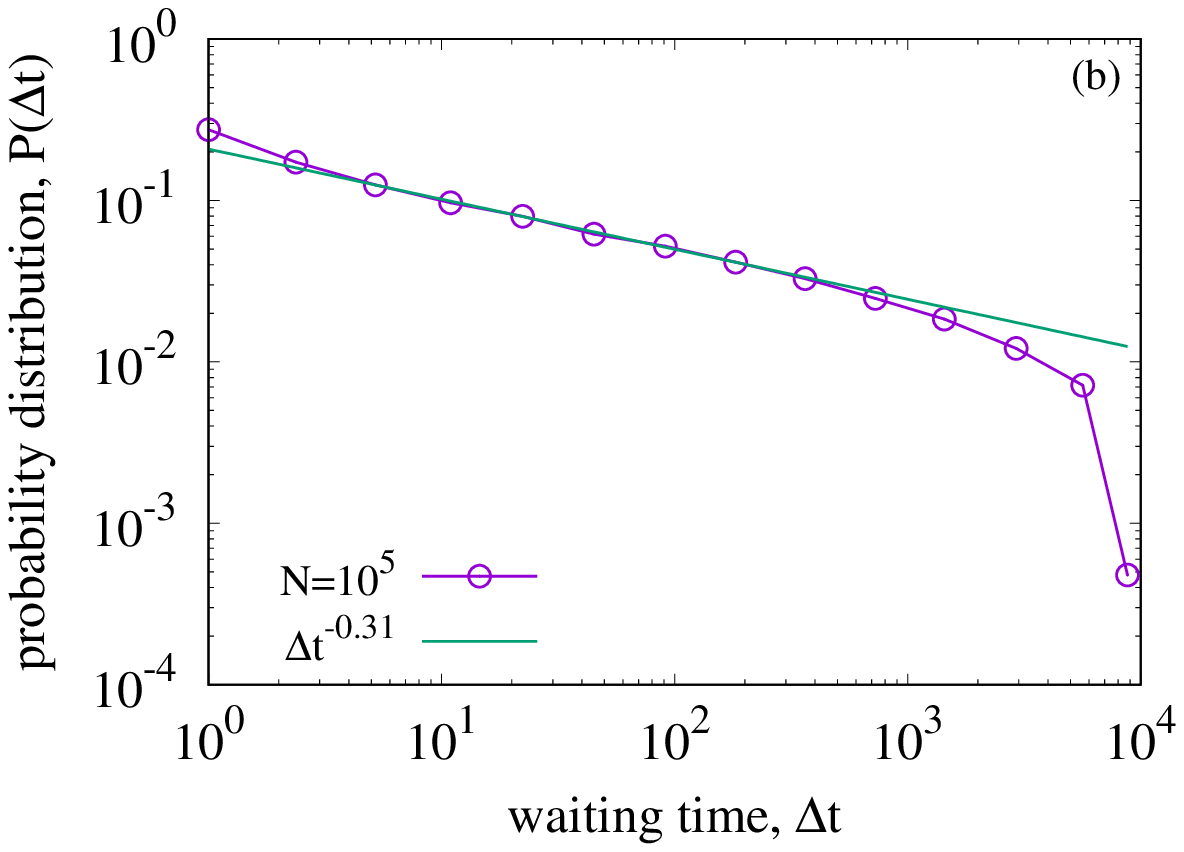}
\caption{The time elapsed between two successive record values of the order parameter is the waiting time $\Delta t$. (a)The probability distribution $P(\Delta t)$ of waiting time $\Delta t$ of velocity record at $F_c=0.3605$ for different system sizes $N=25000, 50000, 100000$ are shown for the EW model. (b)The fitting of $P(\Delta t)$ for large values of $\Delta t$ is shown for $N=10^5$. The distribution $P(\Delta t)$ follows a decaying power law with exponent $\alpha^{EW}_{P} = 0.31 \pm 0.02$.}\label{wait_time_EW}
\end{figure}

Monte Carlo simulation is performed for the EW model of  size $L=10^5$ with periodic boundary condition.
\begin{itemize}

\item At $t=0$, we begin with a flat interface, which we call $h_i(0)=0$ for all $i$. 

\item At each step, at every location $g_i=h_{i+1}(t)+h_{i-1}(t)-2h_i(t)+\eta (i,h)+F_{ext}$ is evaluated, where the first three terms on the right hand side represent the second derivative in Eq. (\ref{ew_eq}), and the
 random pinning force $\eta(i,h)$ is uniformly distributed in (-2,2).

\item The height variables $h_i$ along the interface are updated according to

\begin{eqnarray}
\nonumber h_i(t+1)=
\begin{cases}
h_i(t)+1 & ~~ \text{if}~~ g_i>0\\
h_i(t) & ~~ \text{otherwise}
\end{cases}
\end{eqnarray}

\item The above two steps are repeated at each time.
\end{itemize}
 To find the critical force $F_c$, we calculate the velocity of average height i.e. $V(t)=\frac{d<h>}{dt}$ for different values of driving force $F_{ext}$ as a function of time. When $F_{ext}< F_c$ i.e. when the system is in the pinned state, $V(t)$ decays to zero with time. When $F_{ext}> F_c$ i.e. in the depinned region, $V(t)$  continues to fluctuate around a constant steady state value after about $t=10^5$ when the system comes to a steady state in  the depinned region. The variation of $V(t)$ for different $F_{ext}$ shows that the critical force is $F_c= 0.3605$ (Fig.~\ref{velocity_record}a) for which the velocity decays in a power law.

After achieving the steady state for a particular $F_{ext}>F_c$, we have studied the record statistics for a long time span ($2500000$ time steps), splitting it into $250$ intervals with equal length of $10000$ time steps. To investigate such a statistics we have considered the value of velocity ($V$) at starting instant of each of the time intervals as the first record with record number $n=1$ and have chosen the corresponding time as $t=1$. Thereafter we start comparing the velocity of the subsequent Monte Carlo steps with the recorded maximum velocity and if it is greater than the previous maximum, we count it as the next record and update the last maximum with current velocity value. We continue this process for each of the $250$ time intervals independently and then averaged over those intervals to get time averaged $n(t)$ vs $t$. This has been done for $10$ ensembles and finally averaged over those data. The time and ensemble averaged data of $n(t)$ versus $t$ is shown in 
Fig.~\ref{velocity_record}b.  The curve of $n(t)$ vs $t$ shows that the values of $n(t)$ becomes maximum for $F_{ext}=0.3605$ which is the estimate of the critical force $F_c$ obtained from the time variation of $V(t)$ (Fig.~\ref{velocity_record}a). The density plot in Fig.~\ref{EW_den} shows the variation of the record number with time for different values of the external force. The record number is maximum for the external forces close to the critical force and above the critical force, the record numbers are not as high. The total number of records with time shows a growing power law behavior (see Fig.~\ref{velocity_record}c) in the asymptotic time limit with an exponent value $\alpha^{EW}_{n} = 0.30 \pm 0.01$.

We also measured the waiting time distribution between record events. It is defined as the number of time steps ($\Delta t$) between two successive 
record breaking events.
The probability distributions $P(\Delta t)$ of waiting times $\Delta t$  at the critical point $F_c = 0.3605$ for different system sizes $N=25000, 50000, 100000$ are shown in Fig.~\ref{wait_time_EW}a.  
The distribution function $P(\Delta t)$ decays with $\Delta t$ following a power law with an exponent $\alpha^{EW}_{P} = 0.31 \pm 0.02$ (see Fig.~\ref{wait_time_EW}b).
\begin{figure}[ht]
\centering
\includegraphics[width=8cm]{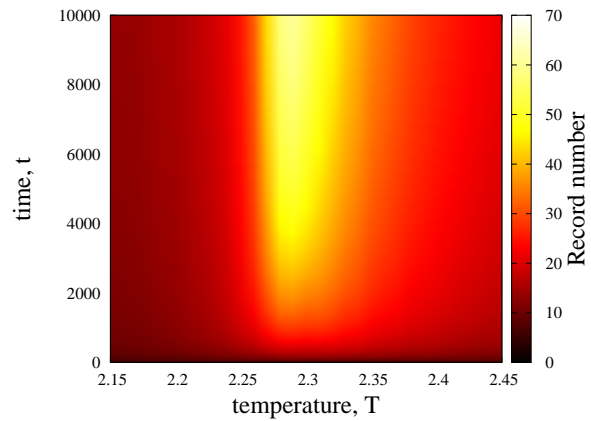}
\caption{The number of record breaking (highest) value for the average magnetization of the Ising model is studied for temperature values both below and above the 
critical value. All measurements are taken after the system reached equilibrium. The plot shows that except for the very initial phase, the number of record breaking events are the highest for 
the temperature near the critical point and the number decreases on both sides of the critical point.}
\label{Ising_den}
\end{figure}
\begin{figure}[ht]
\centering
\includegraphics[width=8cm]{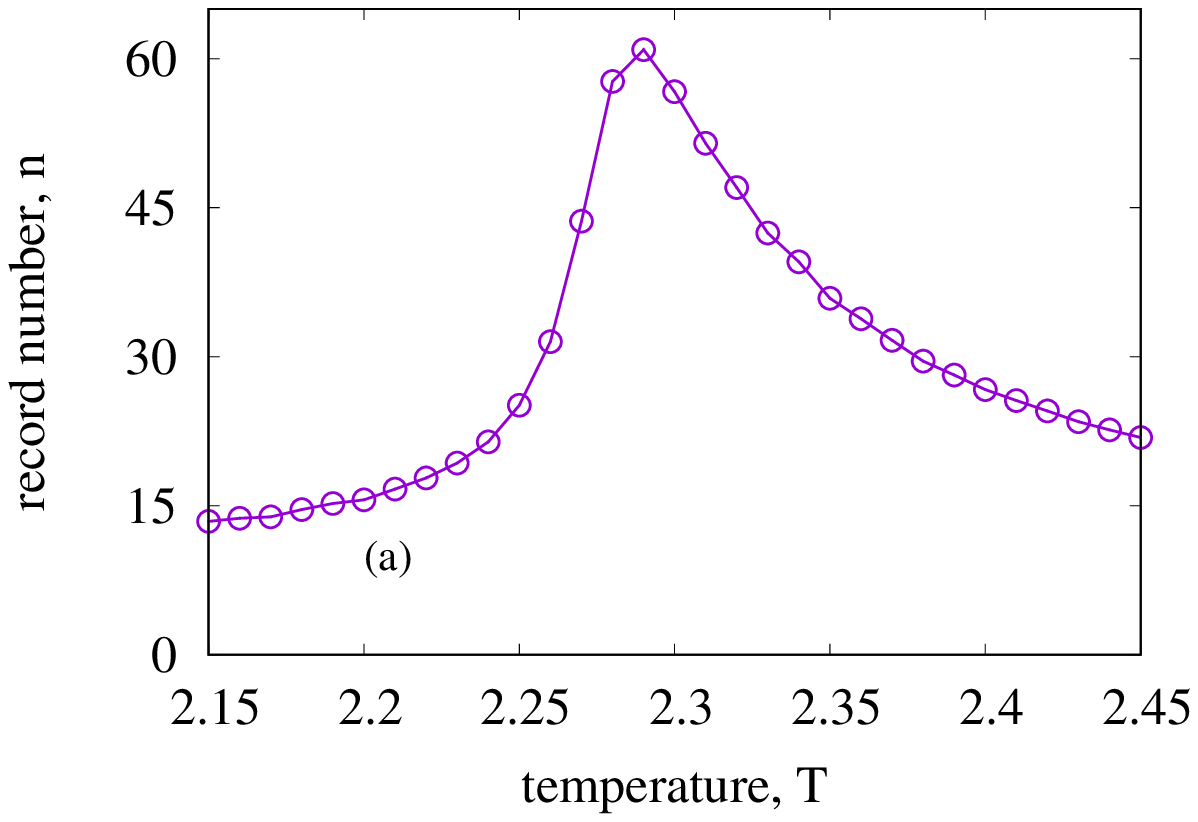}
\includegraphics[width=8cm]{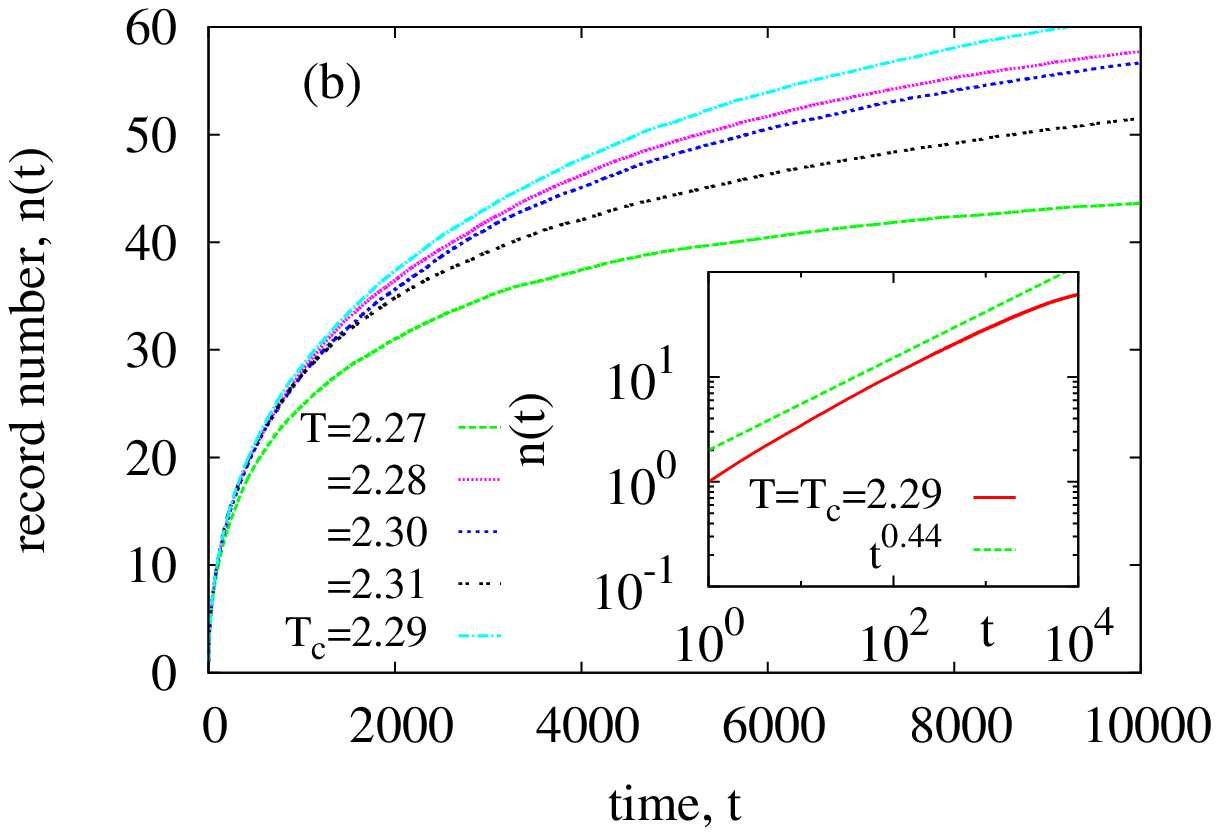}
\caption{The number of record breaking events for the order parameter time series of the two dimensional Ising model is shown for system size $L=200$. (a) Variation of the record number $n$ at $t=10000$ with $T$ is shown.  
 There is a peak of $n$ at $T=2.29$ i.e. the record number is maximum near the critical point. (b) The plots of record number $n(t)$ with time $t$ for different 
 temperatures $T$ are shown. At $T=2.29$ the $n(t)$ vs $t$ curve supersedes the other plots 
 which indicates the critical temperature $T_c = 2.29$. The inset shows, in the limit of large $t$ there is power law rise of $n(t)$ with exponent $\alpha^I_{n} = 0.44\pm 0.01$.}
\label{mag-record}
\end{figure}

\begin{figure}[ht]
\centering
\includegraphics[width=8cm]{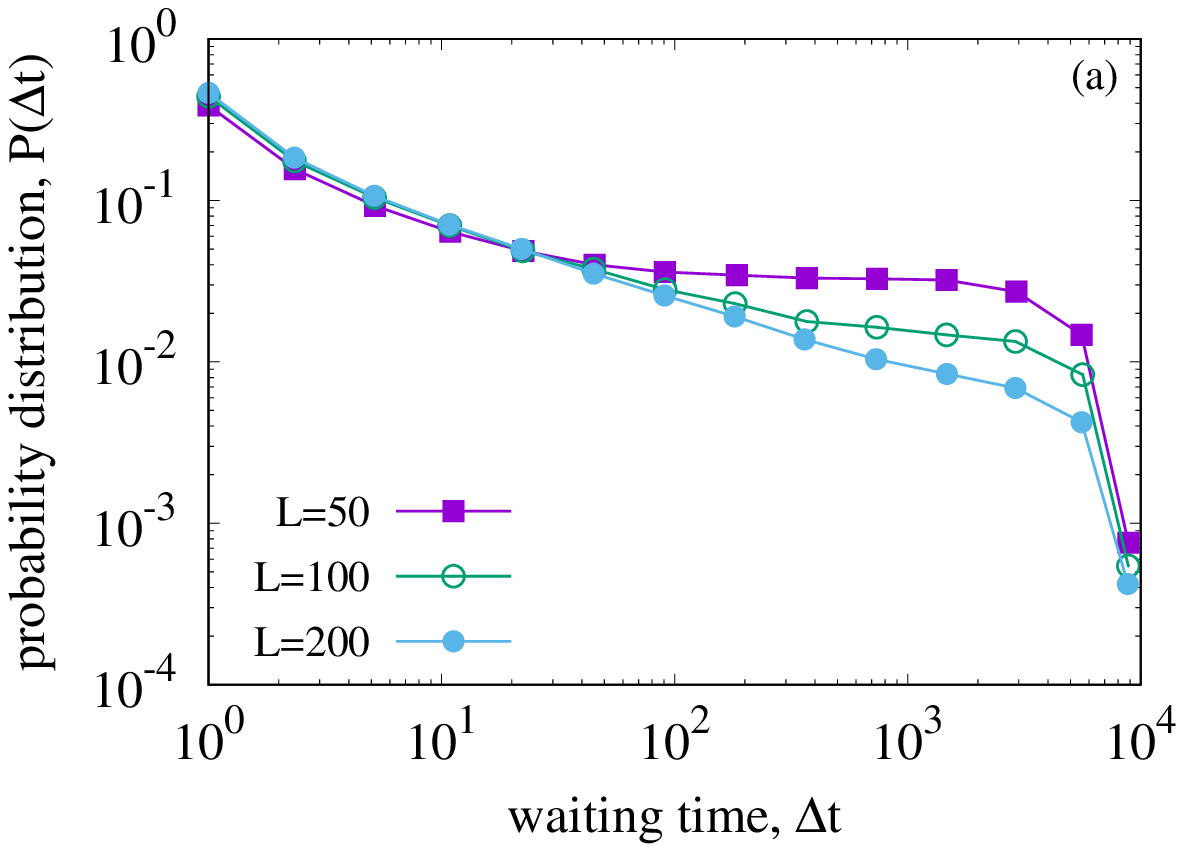}
\includegraphics[width=8cm]{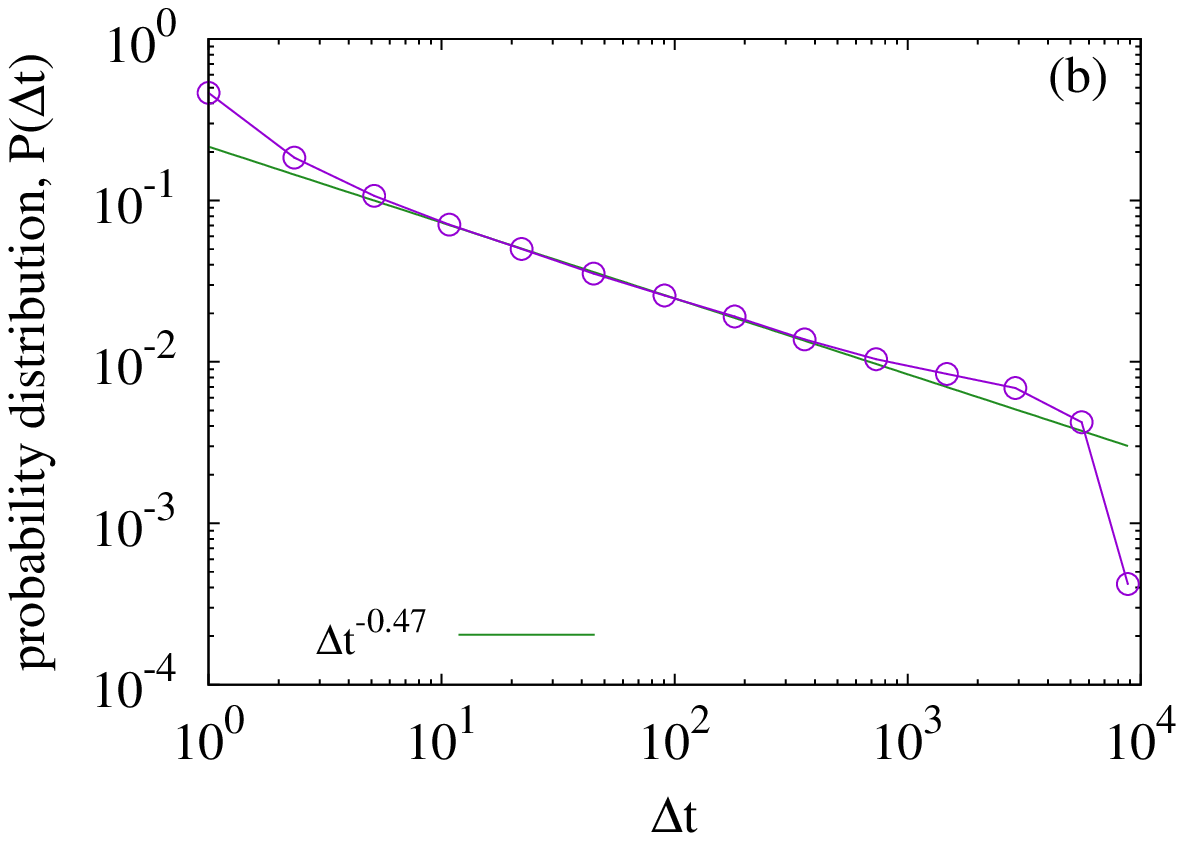}
\caption{(a) The probability distribution $P(\Delta t)$ of waiting time $\Delta t$ between successive record breaking values of 
magnetization at $T_c = 2.29$ for different system sizes $L=50, 100, 200$ are shown. (b) The fitting of 
$P(\Delta t)$ for large values of $\Delta t$ is shown for $L=200$. The distribution $P(\Delta t)$ follows a 
decaying power law with exponent $\alpha^I_{P} = 0.47 \pm 0.02$.}
\label{mag-record-waiting-time}
\end{figure}
\begin{figure}[ht]
\centering
\includegraphics[width=8cm]{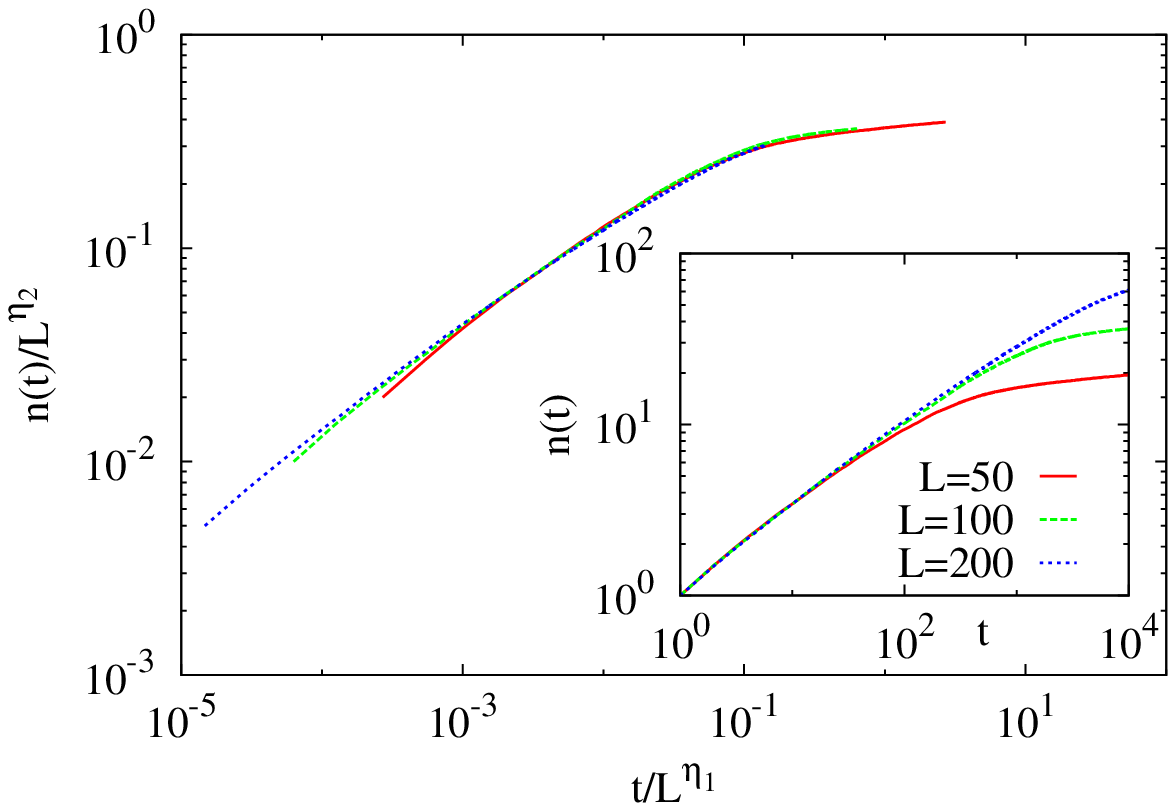}
\caption{The inset shows the number of record events with time for different system sizes near the critical point of two dimensional Ising model. The record number
shows a saturation depending on the system size. The main plot shows the finite size scaling of the data to one curve. The exponent values obtained 
are $\eta_1=2.1$ and $\eta_2=1.0$.}
\label{ising_fss}
\end{figure}
Finally, to check that the critical exponents obtained in the above simulations are universal properties of the model, we change the pinning distributions
from uniform to Gaussian (centered at 0 and having width $\sim$ 3.5). It is found that the exponent values for the number of records with time and for the waiting time
distributions between the records remain unchanged within the numerical accuracies.
Therefore, the record statistics and its associated exponents are characterization of the critical properties of the model. The behavior of the
record statistics can be useful in determining the 
proximity to the critical (depinning) point.

\section{Record statistics of two dimensional Ising Model}
The two dimensional Ising model is a prototypical example of an equilibrium order-disorder phase transition.
First introduced in the context of temperature driven transition in magnetic material \cite{ising}, it has later gained importance
in various fields including binary solids \cite{binder}, neurosciences \cite{hopf}, spin-glass \cite{sudip-binder}, opinion formation \cite{biswas2012disorder,sudip_opinion} in society etc. 
While the one dimensional model does not show a phase transition at any finite temperature, in two dimensions
the critical point is known exactly \cite{onsager}.

Here we focus on the record statistics of the time series of the order parameter near the critical point in the two dimensional Ising model.
While the transition is not associated with a catastrophic failure event,
near critical dynamics in the Ising model can signify sudden and large changes in the sign of the magnitude of the magnetisation. Depending upon
the context, such a switch in polarity can have major consequences (e.g. determination of winner in an election \cite{us_pre})

The Hamiltonian of the two dimensional Ising model $H_{ising}$ of linear size $L$ is
given by (e.g.,~\cite{sudip_stanley})
\begin{align}
H_{ising} = -J\sum_{\langle ij \rangle} S_i^zS_j^z .
\end{align}
Here $S_i^z$ is the $z$-component of spin of the $i$-th site. We consider only the nearest neighbor 
ferromagnetic interaction where $J$ is the strength of the interaction between any pair of spins. Due to the presence 
of ferromagnetic interaction $J$, the spins  try to align along the $z$ direction, which essentially gives 
ferromagnetic magnetization state of the system for a low enough temperature. Such ferromagnetic ordering can be destroyed by increasing the temperature $T$ beyond some critical value $T_c$ where the system becomes paramagnetic.

\begin{figure}[ht]
\centering
\includegraphics[width=8cm]{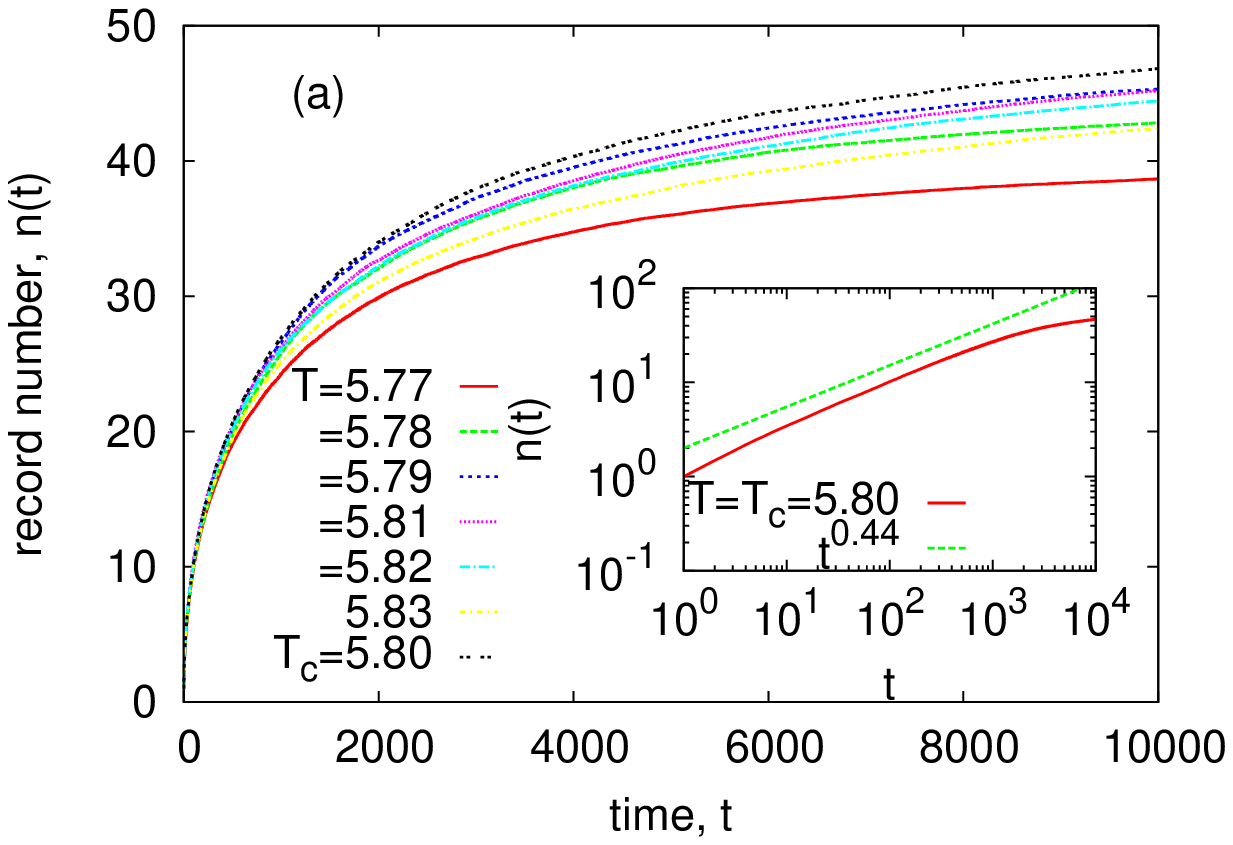}
\includegraphics[width=8cm]{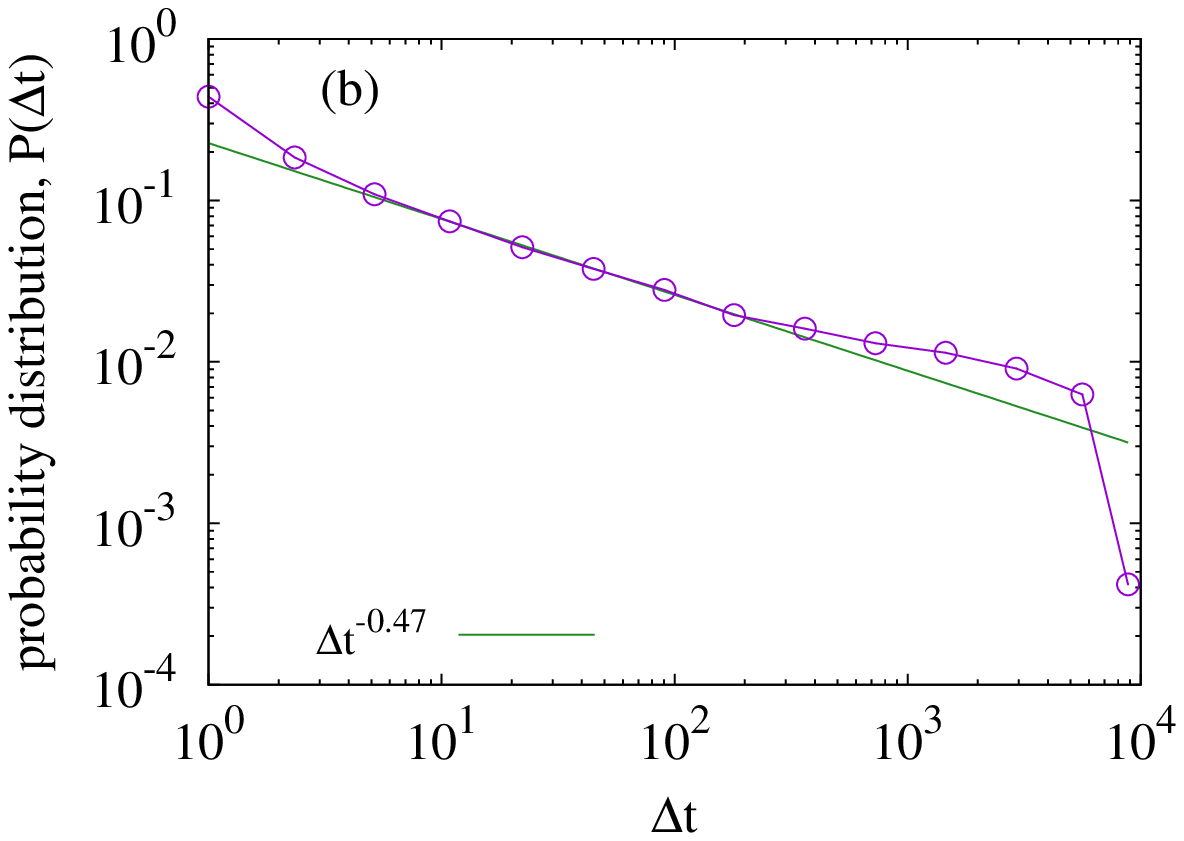}
\caption{(a) The plots of record number $n(t)$ with time $t$ for different 
 temperatures $T$ are shown for the next-nearest neighbor Ising model with system size $L=200$. At $T=5.80$ the $n(t)$ vs $t$ curve supersedes the other plots 
 which indicates the critical temperature $T_c = 5.80$. The inset shows in the limit of large $t$ there is power law rise of $n(t)$ with exponent $\alpha^I_{n} = 0.44 \pm 0.01$. (b) The fitting of $P(\Delta t)$ for large values of $\Delta t$ is shown for the next-nearest neighbor Ising model. The distribution $P(\Delta t)$ follows a 
decaying power law with exponent $\alpha^I_{P} = 0.47 \pm 0.02$.}
\label{mag-record-nnn}
\end{figure}

We perform Monte Carlo simulation on the two dimensional Ising model. 
The average magnetization $m = \frac{1}{L^2} \sum_{i=1}^{L^2} S_i^z$ is  the order 
parameter of the system. We allow the system to equilibrate with $20000$ Monte Carlo steps where 
in one Monte Carlo step every spin is updated just once. To locate the critical point for a finite size, we calculate 
the fluctuation of $m$ which is defined as $\sigma = \overline{\langle m^2 \rangle} - (\overline{\langle m \rangle})^2$. 
Here the overhead bar denotes the configuration average and for such averaging we take $50$ different configurations. 
After equilibration, the thermal averaging $\langle ... \rangle$ is made over $250000$ time steps.

After the equilibration for a given $T$ we study the variation of $m$ with time. As before, for the time averaging, we split the entire time series data into $250$ intervals with equal 
length of $10000$ time steps. In such time series data we fix the starting instant of each of the $250$ intervals as an initial time $t=1$. The corresponding magnitude of $m$ is considered as the first record with number $n=1$. Then walking along the time series we register the successive records when we find the magnitude of $m$ 
to be greater than the previous recorded maximum. We continue such process to extract the variation of $n(t)$ with $t$. The variation of $n(t)$ with $t$ is calculated independently for each of the interval and the time averaging is made over those $250$ intervals and averaged with $10$ ensembles. 
In Fig.~\ref{Ising_den} the number of records increasing with time for different temperatures are shown. In this case both sides of the critical point can be accessed. It is clear that the number of records are lower on either side of the critical point and reaches a maximum near the critical point.
 
\begin{figure}[ht]
\centering
\includegraphics[width=8.5cm,trim={0 2cm 0 0},clip]{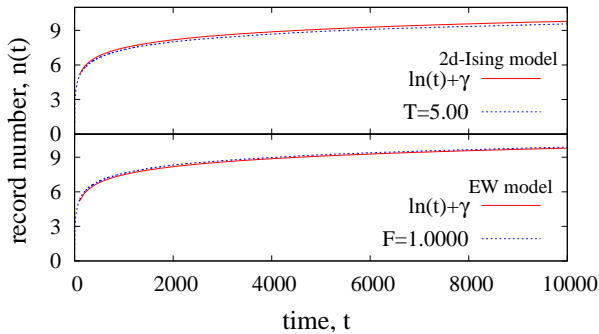}
\caption{Far away from the critical points, the number of record breaking events follow the statistics expected from uncorrelated 
i.i.d. variables. In this plot, we show that for the Ising model at a much higher temperature than the critical point and the EW
model for a much higher external force than the depinning threshold, the number of record breaking events grow logarithmically and 
follow the analytical prediction for i.i.d. variable in the long time limit i.e. $n(t) \to ln(t)+\gamma$, where $\gamma\approx 0.577215 \dots$ is the Euler-Mascheroni constant \cite{rev}.}
\label{ising_iid}
\end{figure}
The time and configuration averaged data of $n$ for several temperatures are shown in Fig.~\ref{mag-record}a. 
  We find that the value of $n$ become maximum for $T=2.29$. This gives the estimation of $T_c$ from the analysis of record statistics which is very close to our previously estimated value. Through the fitting of the 
data of $n(t)$ at $T_c=2.29$, in asymptotic time limit, we find a power law rise of 
$n(t)$ with $t$ and the power law exponent is $\alpha^I_{n} = 0.44 \pm 0.01$ (Fig.~\ref{mag-record}b).  Unlike in the case of the EW model, the record number shows 
a finite size scaling of the form $n(t)=L^{\eta_2}f(t/L^{\eta_1})$, with $\eta_2=1.0$ and $\eta_1=2.1$. The data collapse are shown in Fig.~\ref{ising_fss}.

We compute the probability distribution of waiting time $\Delta t$ at $T_c = 2.29$, where $\Delta t$ is the time interval for getting two consecutive magnetization records. The variation of $P(\Delta t)$ with $\Delta t$ for system sizes $L=50, 100, 200$ are shown in Fig.~\ref{mag-record-waiting-time}a. 
The distribution function $P(\Delta t)$ diminishes with increase of $\Delta t$. We find a power law fall of $P(\Delta t)$ and for system size $L= 200$, best fitting is obtained
with exponent $\alpha^I_{P} = 0.47 \pm 0.02$ (see Fig.~\ref{mag-record-waiting-time}b).

As before, to make sure that the above mentioned exponent values are universal within a given universality class (like all other critical exponents),
 we perform the simulations for the next nearest neighbor Ising model as well. The universality class
of the model is supposed to remain unchanged for this other short range version of the model. The Hamiltonian of the system now has a second (diagonal) neighbor interaction
\begin{equation}
H=-J_1\sum\limits_{NN}S_iS_j-J_2\sum\limits_{NNN}S_iS_j,
\end{equation}
where the second term denotes next nearest neighbor interaction and for simplicity the ratio of the strengths is taken to be $J_1/J_2=1$ (see e.g. \cite{nnn1,nnn2}), where the 
critical temperature is expected to increase from the nearest neighbor model but the critical exponents are expected to remain the same. In Fig. \ref{mag-record-nnn}, the time variation of the record number and 
the waiting time distribution between the records are shown. The exponent values are the same as obtained for the nearest neighbor model within the numerical accuracies. It demonstrates that the exponents associated with the record statistics are characteristics of the universality class, in this case the Ising-class.

Finally, while near the critical points the record numbers  show a power law increase, away from the critical point the growth of the record number with time should be the logarithmic increase  predicted for the i.i.d. statistics \cite{rev}. In Fig.~\ref{ising_iid} the number of record breaking  events for the magnitude of magnetization is plotted for $T=5.0$ for the Ising model, which is far away
from the critical point $T_c$. The variation matches very well (see Fig.~\ref{ising_iid}) with the prediction $n(t)=ln(t)+\gamma$ for the i.i.d. variables. Similarly,  in the case of the EW model, when external force $F_{ext}$ is much higher than the critical force $F_c$, record number follows the same logarithmic behavior (see Fig.~\ref{ising_iid}) expected for temporaly uncorrelated events.

\section{Discussions and conclusion}
Record statistics are the events that has the largest or smallest size of similar kind of events up to that time. While its value, in terms of setting a record, is interesting,
it is also an important tool to understand the temporal clustering of the dynamics, particularly, to understand in any given time series, whether the successive events have a correlation. The number of record breaking events in an uncorrelated time series is known exactly and is independent of the distribution from which the events are drawn. Therefore, the number of record breaking events deviating from the uncorrelated value in a system indicates temporal correlations
developed in the system, which are often associated with fundamental changes in the underlying systems, for example a phase transition.  Furthermore, its detection is free of small scale measurement errors, since by definition a record
breaking event is the largest signal up to that time. Due to these reasons, study of record statistics has gained prominence in 
various fields of science. The strongest fluctuation, and thereby the largest number of record breaking events are, however, likely to
occur near the critical point of a system due to diverging fluctuation. In this work we focused on the 
behavior of record statistics near critical points of some widely used models and associate the critical scaling of various 
quantities to the respective universality classes.

Particularly,
record statistics phenomena is investigated for the EW model and the Ising model around the critical points. Both the models indicate that the asymptotic value of record is maximum at the critical value of the parameter, which drives the related phase transitions. The parameters are transverse force and temperature for EW model and Ising model respectively. The variation of record in long time limit shows power law behavior at the critical point for both the models. It grows with exponent $\alpha^{EW}_{n} \sim 0.30$ for the EW model (Fig. \ref{velocity_record}) and $\alpha^I_{n} \sim 0.44$ for the Ising model (Fig. \ref{mag-record}). The distribution of the waiting time between  records also follow power law behavior. They decay with waiting time having exponents $\alpha^{EW}_{P} \sim 0.31$ (EW model; Fig. \ref{wait_time_EW})  and $\alpha^I_{P} \sim 0.47$ (Ising model; Fig. \ref{mag-record-waiting-time}). 

In conclusion, we have found that the number of record breaking events in a system, within a given time, is maximum near the critical point of that system. 
The number of record breaking events and some associated quantities show power law scaling near the critical point and the exponent values
are identified as characteristics of the respective universality classes of the models. Detections and characterizations of critical points 
in different systems can be done by using the record statistics, which are largely free of small scale detection errors. These results can 
warrant future investigations into more clearly characterizing the relation between critical exponents of the record statistics with other 
critical exponents in the system and using record statistics as precursors to 
imminent catastrophic changes in the system across its critical point. 
\section*{Acknowledgment}

M.K. and S.M. are grateful to Binita Mondal and Sanjukta Paul for their comments and support.  S.B. acknowledges Alexander von Humboldt foundation (Germany) for support during parts of the research period.

\end{document}